\documentclass[aps,prl,twocolumn,groupedaddress,showkeys,showpacs]{revtex4}

\usepackage{multirow,amssymb,amsbsy,amsmath}
\usepackage{epsfig}

\newcommand{\xref}[1]{\protect\ref{#1}}
\newcommand{\figref}[1]{Fig.~\protect\ref{#1}}

\newcommand{\Bcrit}{B_{\text{c}}}
\newcommand{\Bsat}{B_{\text{sat}}}
\newcommand {\mofe} {\{$\textrm{Mo}_{72}\textrm{Fe}_{30}$\}}

\begin{document}

\title{Metamagnetic phase transition of the antiferromagnetic Heisenberg icosahedron}

\author{Christian Schr\"oder}
\email{christian.schroeder@fh-bielefeld.de}
\affiliation{Department of Electrical Engineering and Computer
  Science, University of Applied Sciences Bielefeld, D-33602 Bielefeld, Germany
  \& Ames Laboratory, Ames, Iowa 50011, USA}

\author{Heinz-J\"urgen Schmidt}
\author{J\"urgen Schnack}
\affiliation{Universit\"at Osnabr\"uck, Fachbereich Physik,
D-49069 Osnabr\"uck, Germany}

\author{Marshall Luban}
\affiliation{Ames Laboratory \& Department of Physics and Astronomy,
Iowa State University, Ames, Iowa 50011, USA}

\date{\today}

\begin{abstract}
  The observation of hysteresis effects in single molecule
  magnets like Mn$_{12}$-acetate has initiated ideas of future
  applications in storage technology.  The appearance of a
  hysteresis loop in such compounds is an outcome of their
  magnetic anisotropy.  In this Letter we report that magnetic
  hysteresis occurs in a spin system without any anisotropy,
  specifically, where spins mounted on the vertices of an
  icosahedron are coupled by antiferromagnetic isotropic
  nearest-neighbor Heisenberg interaction giving rise to
  geometric frustration.  At $T=0$ this system undergoes a first
  order metamagnetic phase transition at a critical field
  $\Bcrit$ between two distinct families of ground state
  configurations. The metastable phase of the system is
  characterized by a temperature and field dependent survival
  probability distribution.
\end{abstract}

\pacs{75.50.Xx;75.10.Hk;75.10.Jm;75.30.Cr}
\keywords{Heisenberg model, Metamagnetic Phase Transition}
\maketitle

%%%%%%%%%%%%%%%%%%%%%%%%%%%%%%%%%%%%%%%%%%%%%%%%%%%%%%%%%%%%%%%%%%%%%%%%

\begin{figure}[ht!]
\centering
\includegraphics[clip,width=65mm]{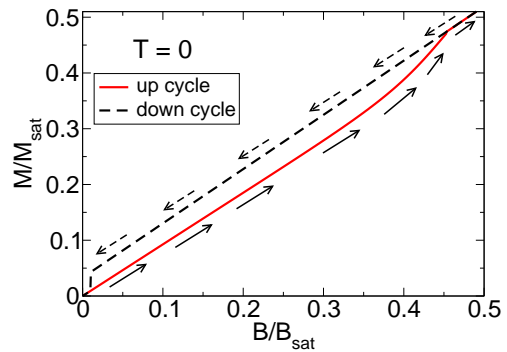}
\caption{Hysteresis behavior of the classical icosahedron
   in an applied magnetic field
   obtained by classical spin dynamics simulations.}
\label{icosa_jump}
\end{figure}

\emph{Introduction}---Low-dimensional magnetic systems show a
variety of fascinating phenomena that are associated with
geometrical frustration \cite{Gre:JMC01,SRF:LNP04}.  Among them
are magnetization plateaus and jumps as well as unusual
susceptibility minima, as observed for example for the kagome
lattice antiferromagnet \cite{NKH:EPL04,SHS:PRL02}.  Some of
these effects can also occur in certain strongly frustrated
magnetic molecules such as the Keplerate \mofe \cite{SNS:PRL05}.
In this Letter we report that a first order metamagnetic phase
transition (with associated hysteresis and metastability
effects) occurs for a system of spins that are mounted on the
vertices of an icosahedron when an external magnetic field, $B$,
equals a critical value $\Bcrit$.  These spins interact with one
another only via nearest-neighbor, antiferromagnetic isotropic
Heisenberg exchange, but due to the geometrical frustration
originating in the unique geometry of the icosahedron they
undergo the metamagnetic transition despite the absence of any
anisotropic energy terms.  As the field proceeds through a closed
cycle the magnetization traces the hysteresis loop shown in
\figref{icosa_jump}.  The metastable phases, including the
temperature dependence of their lifetime, show rich
characteristics.

%===================    figure   =================================
%===================    figure   =================================

Our exact classical treatment shows that the abrupt transition at
$T=0$ originates in the intersection of two energy curves
belonging to different families of spin configurations that are
ground states below and above the critical field.  The minimum
of the two energy functions constitutes a non-convex minimal
energy function of the spin system and this gives rise to a
metamagnetic phase transition \cite{LhM:02}. At $T=0$ the
partition function is a non-analytic function of $B$, and since
the magnetization features a finite jump at $\Bcrit$ the
transition is of first order.  We also show that the
corresponding quantum spin system for sufficiently large spin
quantum number $s$ possesses a non-convex set of lowest energy
levels when plotted versus total spin. This is the discrete
analog of the non-convex classical minimal energy function.
Therefore, also the quantum spin system features an unusual
magnetization jump.

%%%%%%%%%%%%%%%%%%%%%%%%%%%%%%%%%%%%%%%%%%%%%%%%%%%%%%%%%%%%%%%%%%%%%%%%%%%%%%%%%%%%%%%%%%%%%%%%%
\emph{Classical simulations at $T=0$}---The behavior of
classical spin systems subject to an applied magnetic field both
at $T\approx 0$ and finite temperatures can be very effectively
studied with the help of a stochastic spin dynamics approach
such as that proposed in \cite{ATH:JAP97}. Here, the spin system
is coupled to a heat bath in a Langevin-type approach by using a
Landau-Lifshitz-damping term as well as a fluctuating force with
white noise characteristics.  Starting from an arbitrary initial
configuration the spin system can be investigated either at zero
temperature by observing the relaxation to its ground state or
at finite temperature by following its time evolution.

We consider first $T=0$. For $B=0$ we find, that the spins adopt
a non-coplanar configuration, where each of the 12 spin vectors
makes an angle of $\text{arccos}(-1/\sqrt{5})\approx
116.6^\circ$ with respect to its five nearest neighbors.  This
is in agreement with the analytical results derived in
\cite{ScL:JPA03}.

The spins are also subjected to an external magnetic field that
increases from zero linearly with time but at a very slow rate.
The dynamical evolution of the spins is monitored as a function
of time and the results are stored within an animation file
\footnote{We invite the reader to examine movies of our
  simulations at http://www.fh-bielefeld.de/fb2/schroeder}
enabling direct visualization of the spin vectors.  In the first
stage the configuration of the spin vectors evolves continuously
with $B$ until suddenly there is an abrupt change in their
orientations.  In the new configuration two spin vectors at
opposite ends of a diameter of the icosahedron are aligned
parallel to the field while all of the remaining 10 spin vectors
subtend a common polar angle with respect to $\vec{B}$ and their
azimuthal angles are uniformly spaced.  With increasing $B$ the
common polar angle decreases monotonically while the azimuthal
angles remain fixed until at saturation all spin vectors are
parallel to $\vec{B}$.

%%%%%%%%%%%%%%%%%%%%%%%%%%%%%%%%%%%%%%%%%%%%%%%%%%%%%%%%%%%%%%%%%%%%%%%%%%%%%%%%%%%%%%%%%%%%%%%%%
\emph{Analytical results}---For a classical
spin system the ground states are defined as states
$\vec{s}=(\vec{s}(1), \ldots, \vec{s}(N))$ which minimize the
energy
%--------------------------------------------------------
\begin{eqnarray}
\label{TF1}
H(\vec{s}, B)
&=&
H_0(\vec{s})
-
B
\sum_{\mu} s_z(\mu)
\\
&=&
\sum_{\mu,\nu}\;J_{\mu\nu}
\vec{s}(\mu)\cdot\vec{s}(\nu)
-
B
\sum_{\mu} s_z(\mu)
\nonumber
\ .
\end{eqnarray}
%--------------------------------------------------------
The coupling $J_{\mu\nu}$ between spins $\mu,\nu$ is chosen as
$J=1$ between nearest-neighbors and zero otherwise.  Accordingly
the magnetic field is given in appropriate units.

The (degenerate) ground states must fulfill the following
necessary condition, compare Eq.~(21) in \cite{ScL:JPA03},
%--------------------------------------------------------
\begin{equation}
\label{TF2}
\sum_{\nu}J_{\mu\nu}\vec{s}(\nu)
=
\kappa_\mu \vec{s}(\mu)
+
\frac{1}{2}\vec{B}
\quad (\mu=1,\ldots,N)
\ ,
\end{equation}
%--------------------------------------------------------
where $\kappa_\mu$ denote suitable Lagrange parameters.
Although this system of equations can only be solved numerically
in most cases, we are confident that the following statements
about ground states of the icosahedron subject to magnetic
fields are correct \cite{SSS:05}.

As remarked above, the spin configuration for $B=0$ has been
established \cite{ScL:JPA03}.  The orientation of the individual
spins is such that they form four groups of three spins where
each group $i$ is characterized by a common polar angle
$\theta_i$ and uniformly spaced azimuthal angles.

For $B>0$ we numerically solve (\ref{TF2}) with the assumption
that the azimuthal angles remain fixed and only the four polar
angles vary \footnote{This is confirmed by numerical methods
  \cite{SSS:05}.}.  Thus we obtain a $1$-parameter family (the
``$4$-$\theta$-family'') of possible ground states.
Interestingly, it provides a local minimum of the energy only
for $0\le M\le 5.61441$.  One might assume that the
$4$-$\theta$-family also provides a global minimum of energy,
i.e.~a ground state, for the same interval $0\le M\le 5.61441$,
but this is wrong: There exists a different $1$-parameter family
of solutions of (\ref{TF2}) which has a lower energy than the
$4$-$\theta$-family for $M>M_0\simeq 4.92949$. As remarked
above, this family can be characterized by two spin vectors
which are aligned parallel to $\vec{B}$ and 10 spin vectors with a
common polar angle $\theta$ and uniformly spaced relative
azimuthal angles. The end points of the latter 10 spin vectors
form a regular decagon, and we will call this set of states the
``decagon family''.  As discussed below, the decagon family
provides a local minimum of the energy for $0.54102\le M \le
12$. These families give rise to two convex curves in the $E$
vs.~$M$ diagram, $E_1(M)$ for the $4$-$\theta$-family and
$E_2(M)$ for the decagon-family, which intersect at the point
with coordinates $(M_0,E_0)$, where $M_0\simeq 4.92949$ and
$E_0\simeq -11.150$.

%===================    figure   =================================
\begin{figure}[ht!]
\centering
\includegraphics[clip,height=40mm]{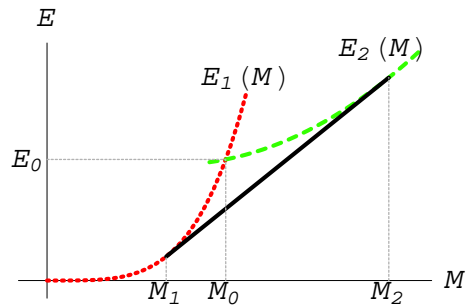}
\caption{Schematic representation of two minimal energy curves
  $E_1(M)$ and $E_2(M)$, whose minimum $E_{\text{min}}(M)$ is
  not convex. Their common tangent (solid line) has a slope of
  $\Bcrit$.} 
\label{decagon-ct}
\end{figure}
%===================    figure   =================================

We have strong numerical evidence that the minimum of the two
curves provides the absolute minimum $E_{\mbox{\scriptsize
    min}}(M)$ of $H_0(\vec{s})$ for given $M$ \cite{SSS:05}.
The latter function $E_{\mbox{\scriptsize min}}(M)$ is therefore
not convex and this translates to a jump in the magnetization,
$\Delta M=M_2-M_1$, at a critical field $\Bcrit$, compare
\cite{LhM:02}. One can identify $\Bcrit$ as the slope of the
common tangent of the curves $E_1(M)$ and $E_2(M)$ illustrated
in \figref{decagon-ct}.  This construction is fully equivalent
to the statement that the total energies of the two phases,
$E_1(M_1)- B M_1$ and $E_2(M_2)- B M_2$, are equal for
$B=\Bcrit$.  According to the Ehrenfest classification the phase
transition is of first order.  Equivalently, $\Bcrit$ can be
obtained by a Maxwell construction in the $M$ vs.~$B$ diagram.
The pertinent quantities for the phase transition are given in
Table \xref{Table1}.

%--------------------------------------------------------
\begin{table}
  \caption{Characteristic values of the first-order phase
    transition of the spin icosahedron, compare
    \figref{decagon-ct}. 
    $\Bsat=2(5+\sqrt{5})$ denotes the saturation field.
    $\chi_i=\frac{dM_i}{dB}|_{B=\Bcrit},\, i=1,2$ denote the
    limit values of the susceptibility. 
  }
\label{Table1}
\begin{center}
\begin{tabular}{ll}
$\Bcrit\simeq 5.87614$              & $\Bcrit\simeq 0.40603\ \Bsat$\\
$M_{1}\simeq 4.71461$               & $M_{2}\simeq 5.10784$\\
$E_1(M_1)\simeq -12.4324\quad$           & $E_2(M_2)\simeq -10.1218$\\
$E_0\simeq -11.150$                 & $ M_0 \simeq 4.92949$\\
$\chi_{1}\simeq 1.13294$            & $\chi_{2}=\frac{5}{4+\sqrt{5}}\simeq 0.8018$
\end{tabular}
\end{center}
\end{table}
%--------------------------------------------------------

Since $M(B)$ has a jump at $B=\Bcrit$ the susceptibility $\chi=
\frac{dM}{dB}$ diverges at $B=\Bcrit$ and $T=0$. It is rather
easy to derive the leading temperature dependence of this
divergence for $T\rightarrow 0$. Recall that an alternative
expression for the susceptibility is
%--------------------------------------------------------
\begin{equation}
\label{CE1}
\chi(B,\beta)
=
\beta\left(\langle M^2 \rangle_\rho - \langle M \rangle_\rho^2
\right)
\equiv
\beta \sigma^2(M)
\,,
\end{equation}
%--------------------------------------------------------
where $\langle\quad\rangle_\rho$ denotes the thermal expectation
value w.~r.~t.~the canonical ensemble $\rho=\rho(\beta,B)$.  In
the case of a metamagnetic phase transition of the kind
described above the ground state is degenerate for $T\rightarrow
0$ and $B\rightarrow \Bcrit$. Hence $\sigma^2(M)$ remains finite
in this limit and $\chi(\Bcrit,\beta)$ diverges linearly with
$\beta$, i.e. the critical exponent is 1.

%%%%%%%%%%%%%%%%%%%%%%%%%%%%%%%%%%%%%%%%%%%%%%%%%%%%%%%%%%%%%%%%%%%%%%%%%%%%%%%%%%%%%%%%%%%%%%%%%
\emph{Stability and hysteresis}---In order to investigate the
stability of the two families of ground states we performed a
standard stability analysis by constructing the Hesse matrix,
which for this system is a symmetric $24\times 24$ matrix.
Since the energy is invariant under rotations about the
$z$-axis, one eigenvalue must be zero. We call a state
satisfying (\ref{TF2}) ``stable" if the Hesse matrix has only
positive eigenvalues apart from one zero eigenvalue. This is
equivalent to the fact that the given state provides a local
minimum of the energy.

By applying this procedure to the two families of possible
ground states of the icosahedron we determine numerically the
above-mentioned stability ranges of the families: The
$4$-$\theta$-family is stable for $0\le M\le 5.61441$ and the
decagon family is stable for $0.54102\le M \le 12$.

This implies that the system at $T\simeq 0$ will not immediately
jump from the $4$-$\theta$-family into the decagon family if $B$
increases beyond $\Bcrit$ but remain in its family until
$M>5.61441$. Conversely, the decagon family will remain the
\emph{de facto} spin configuration of the icosahedron if $B$ is
lowered beyond $\Bcrit$ until $M<0.54102$. In fact, these
hysteresis effects are observed in our simulational studies, see
\figref{icosa_jump}.

%%%%%%%%%%%%%%%%%%%%%%%%%%%%%%%%%%%%%%%%%%%%%%%%%%%%%%%%%%%%%%%%%%%%%%%%%%%%%%%%%%%%%%%%%%%%%%%%%
\emph{Quantum calculations}---We now discuss how the
metamagnetic phase transition is manifested in the quantum
Heisenberg icosahedron. Classically, the phase transition
consists of a discontinuity of the magnetization as a function
of the magnetic field. Quantum mechanically, the magnetization
curve for $T=0$ is already a staircase of successive steps of
unit height $\Delta M = 1$, which result from crossings of
levels with adjacent total magnetic quantum numbers $M$ and
$M+1$.  In the context of this phase transition we are looking
for a magnetization jump of unusual height, i.e.  $\Delta M >
1$. It is clear that such a jump must occur because it occurs in
the classical limit $s\rightarrow\infty$. The remaining
questions therefore are, for which intrinsic spin quantum number
$s$ is such a jump clearly visible, and are there other signs of
the phase transition at smaller $s$?

%===================    figure   =================================
\begin{figure}[ht!]
\centering
\includegraphics[clip,width=55mm]{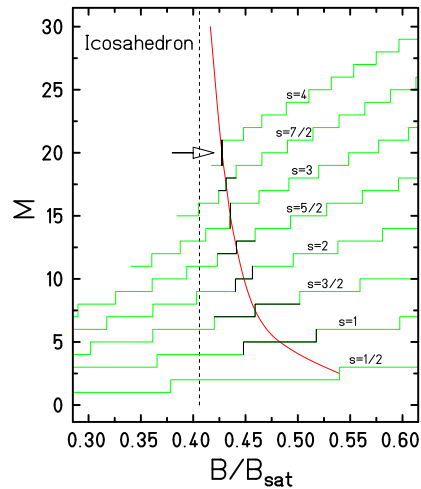}
\caption{Magnetization curves for $T=0$ and various values of
  the intrinsic spin quantum number $s$: The magnetization
  plateaus of smallest width are highlighted on each curve. At
  $s=4$ a magnetization jump of $\Delta M = 2$ occurs, 
  marked by the arrow. The solid curve shows that the field
  values that bisect the smallest plateaus converge to the
  classical transition field (dashed line).  For large $s$ the
  magnetization curve could only be evaluated down to a certain
  field due to prohibitively large dimensions of the related
  Hilbert spaces.}
\label{icosahedron-M-B-all}
\end{figure}
%===================    figure   =================================

Using a Lanczos procedure which yields numerically exact lowest
energy eigenvalues in subspaces with constant total magnetic
quantum number $M$ we are able to evaluate magnetization curves
at $T=0$ for various intrinsic spin quantum numbers.  Figure
\ref{icosahedron-M-B-all} shows the relevant parts of the
magnetization curves at $T=0$ for $s=1/2,\dots,4$.  For integer
values of $s$ the magnetization plateau of smallest width is
highlighted; for half-integer values of $s$ we highlight two
such plateaus. With increasing $s$ these widths shrink and
already at $s=4$ a magnetization jump of $\Delta M = 2$ occurs.
This corresponds to a non-convex part of the discrete energy
levels versus $M$.

In \figref{icosahedron-M-B-all} we also provide a curve which
bisects the magnetization plateaus of smallest width. Assuming
that the bisector value of $B/\Bsat$ is described by a
polynomial in $1/s$, we obtain as an estimate for the classic
transition field $B/\Bcrit\approx 0.40\pm0.01$, which is in very
good agreement with the classical result (see
Table~\xref{Table1}).  The uncertainty originates from the
limited number of data points (eight) as well as from
fluctuations between integer and half-integer values of $s$.

%%%%%%%%%%%%%%%%%%%%%%%%%%%%%%%%%%%%%%%%%%%%%%%%%%%%%%%%%%%%%%%%%%%%%%%%%%%%%%%%%%%%%%%%%%%%%%%%%
\emph{Classical finite-temperature simulations}---We have
restricted our investigation to the determination of the
lifetime of the high-field phase (decagon-family) in its
metastable regime ($B/\Bcrit<1$).  The lifetime has been
determined by the following procedure: First, the system is
prepared in the decagon phase at $T=0$. Then the field is
lowered to a value below the critical field value $\Bcrit$.
Starting from these initial conditions the temperature is set to
a value $T>0$ and the trajectory of the system is calculated
numerically by solving the stochastic Landau-Lifshitz equation.

\begin{figure}[ht!]
\centering
\includegraphics[clip,width=60mm]{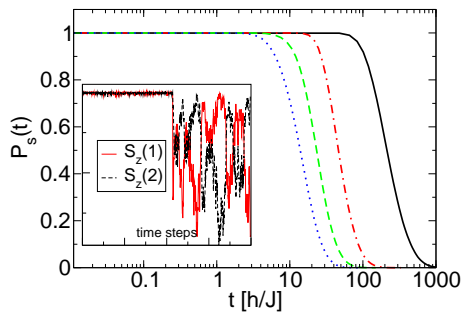}
\caption{Survival probability $P_{\text{s}}(t)$
for the metastable decagon phase subject to
an external field $B/\Bsat=0.27$
for temperatures $k_B T/J=0.025, 0.015, 0.005, 0.0005$
(left to right).
Inset: Example trajectory of the system's time evolution at
  finite temperature.
}
\label{survival}
\end{figure}

Shown in the inset of \figref{survival} is a single trajectory
of a sample system.  We exploit the unique property of the
decagon family that two spins persist in pointing in the
direction of $\vec{B}$ ($s_z(1)\approx s_z(2)\approx 1$) until
abruptly breaking away. This allows one to obtain an accurate
determination of the decay time for this system.  By performing
$10^5$ such runs for each choice of $T$ and analyzing the
histograms of the resulting decay times one can determine the
lifetime distribution. A common measure is the so-called
survival probability $P_{\text{s}}(t)$ which is the probability
of the metastable state not having decayed by the time $t$.  In
\figref{survival} we have plotted $P_{\text{s}}(t)$ for the
metastable state for various temperatures and an external field
in the metastable regime.  An appropriate choice for the
lifetime, $t_s$, is the root of $P_s(t_s)=0.5$.  We find that
$t_s$ increases with decreasing temperature and appears to
diverge for $T\rightarrow 0$ as $1/T$.  Although one obtains
similar probability distributions for systems showing thermally
activated magnetization switching \cite{BNR:JAP00}, we emphasize
that our model Hamiltonian does not contain any additional
energy term providing an energy barrier. In fact, it is the
special geometry of the icosahedron that causes the system to
show metastability.

%===================    figure   =================================
%===================    figure   =================================

%%%%%%%%%%%%%%%%%%%%%%%%%%%%%%%%%%%%%%%%%%%%%%%%%%%%%%%%%%%%%%%%%%%%%%%%%%%%%%%%%%%%%%%%%%%%%%%%%
\emph{Summary and outlook}---In this Letter we have shown that
the antiferromagnetic Heisenberg spin icosahedron undergoes a
metamagnetic phase transition and displays rich hysteresis and
metastability phenomena when subject to a varying external
field.  Given this, the metamagnetic transition of the
Heisenberg icosahedron may be of interest for potential
applications in the area of nanomagnetic switches. It is also
relevant in connection with magnetocalorics, since the
magnetization jump is accompanied by an enhanced magnetocaloric
effect \cite{DeR:PRB04,ZhH:JSM04}.  It is therefore very
encouraging that recent progress in the synthesis of magnetic
molecules offers the prospect of realizing the Heisenberg
icosahedron \cite{BGG:JCSDT97,BGH:JCSDT97}.

%%%%%%%%%%%%%%%%%%%%%%%%%%%%%%%%%%%%%%%%%%%%%%%%%%%%%%%%%%%%%%%%%%%%%%%%%%%%%%%%%%%%%%%%%%%%%%%%%
\emph{Acknowledgment}---We thank Peter Hage for the use of his
Lanczos diagonalization routine and R.E.P. Winpenny for
discussions on the possible synthesis of a Heisenberg
icosahedron. Ames Laboratory is operated for the U.S. Department
of Energy by Iowa State University under Contract No.
W-7405-Eng-82.

%\bibliography{js-icosa}

\begin{thebibliography}{14}
\expandafter\ifx\csname natexlab\endcsname\relax\def\natexlab#1{#1}\fi
\expandafter\ifx\csname bibnamefont\endcsname\relax
  \def\bibnamefont#1{#1}\fi
\expandafter\ifx\csname bibfnamefont\endcsname\relax
  \def\bibfnamefont#1{#1}\fi
\expandafter\ifx\csname citenamefont\endcsname\relax
  \def\citenamefont#1{#1}\fi
\expandafter\ifx\csname url\endcsname\relax
  \def\url#1{\texttt{#1}}\fi
\expandafter\ifx\csname urlprefix\endcsname\relax\def\urlprefix{URL }\fi
\providecommand{\bibinfo}[2]{#2}
\providecommand{\eprint}[2][]{\url{#2}}

\bibitem[{\citenamefont{Greedan}(2001)}]{Gre:JMC01}
\bibinfo{author}{\bibfnamefont{J.}~\bibnamefont{Greedan}}, \bibinfo{journal}{J.
  Mater. Chem.} \textbf{\bibinfo{volume}{11}}, \bibinfo{pages}{37}
  (\bibinfo{year}{2001}).

\bibitem[{\citenamefont{Schollw\"ock et~al.}(2004)\citenamefont{Schollw\"ock,
  Richter, Farnell, and Bishop}}]{SRF:LNP04}
\bibinfo{editor}{\bibfnamefont{U.}~\bibnamefont{Schollw\"ock}},
  \bibinfo{editor}{\bibfnamefont{J.}~\bibnamefont{Richter}},
  \bibinfo{editor}{\bibfnamefont{D.}~\bibnamefont{Farnell}}, \bibnamefont{and}
  \bibinfo{editor}{\bibfnamefont{R.}~\bibnamefont{Bishop}}, eds.,
  \emph{\bibinfo{title}{Quantum Magnetism}}, vol. \bibinfo{volume}{645} of
  \emph{\bibinfo{series}{Lecture Notes in Physics}}
  (\bibinfo{publisher}{Springer}, \bibinfo{address}{Berlin, Heidelberg},
  \bibinfo{year}{2004}).

\bibitem[{\citenamefont{Narumi et~al.}(2004)\citenamefont{Narumi, Katsumata,
  Honda, Domenge, Sindzingre, Lhuillier, Shimaoka, Kobayashi, and
  Kindo}}]{NKH:EPL04}
\bibinfo{author}{\bibfnamefont{Y.}~\bibnamefont{Narumi}},
  \bibinfo{author}{\bibfnamefont{K.}~\bibnamefont{Katsumata}},
  \bibinfo{author}{\bibfnamefont{Z.}~\bibnamefont{Honda}},
  \bibinfo{author}{\bibfnamefont{J.-C.} \bibnamefont{Domenge}},
  \bibinfo{author}{\bibfnamefont{P.}~\bibnamefont{Sindzingre}},
  \bibinfo{author}{\bibfnamefont{C.}~\bibnamefont{Lhuillier}},
  \bibinfo{author}{\bibfnamefont{Y.}~\bibnamefont{Shimaoka}},
  \bibinfo{author}{\bibfnamefont{T.~C.} \bibnamefont{Kobayashi}},
  \bibnamefont{and} \bibinfo{author}{\bibfnamefont{K.}~\bibnamefont{Kindo}},
  \bibinfo{journal}{Europhys. Lett.} \textbf{\bibinfo{volume}{65}},
  \bibinfo{pages}{705} (\bibinfo{year}{2004}).

\bibitem[{\citenamefont{Schulenburg et~al.}(2002)\citenamefont{Schulenburg,
  Honecker, Schnack, Richter, and Schmidt}}]{SHS:PRL02}
\bibinfo{author}{\bibfnamefont{J.}~\bibnamefont{Schulenburg}},
  \bibinfo{author}{\bibfnamefont{A.}~\bibnamefont{Honecker}},
  \bibinfo{author}{\bibfnamefont{J.}~\bibnamefont{Schnack}},
  \bibinfo{author}{\bibfnamefont{J.}~\bibnamefont{Richter}}, \bibnamefont{and}
  \bibinfo{author}{\bibfnamefont{H.-J.} \bibnamefont{Schmidt}},
  \bibinfo{journal}{Phys. Rev. Lett.} \textbf{\bibinfo{volume}{88}},
  \bibinfo{pages}{167207} (\bibinfo{year}{2002}).

\bibitem[{\citenamefont{Schr{\"o}der et~al.}(2005)\citenamefont{Schr{\"o}der,
  Nojiri, Schnack, Hage, Luban, and K{\"o}gerler}}]{SNS:PRL05}
\bibinfo{author}{\bibfnamefont{C.}~\bibnamefont{Schr{\"o}der}},
  \bibinfo{author}{\bibfnamefont{H.}~\bibnamefont{Nojiri}},
  \bibinfo{author}{\bibfnamefont{J.}~\bibnamefont{Schnack}},
  \bibinfo{author}{\bibfnamefont{P.}~\bibnamefont{Hage}},
  \bibinfo{author}{\bibfnamefont{M.}~\bibnamefont{Luban}}, \bibnamefont{and}
  \bibinfo{author}{\bibfnamefont{P.}~\bibnamefont{K{\"o}gerler}},
  \bibinfo{journal}{Phys. Rev. Lett.} \textbf{\bibinfo{volume}{94}},
  \bibinfo{pages}{017205} (\bibinfo{year}{2005}).

\bibitem[{\citenamefont{Lhuillier and Misguich}(2002)}]{LhM:02}
\bibinfo{author}{\bibfnamefont{C.}~\bibnamefont{Lhuillier}} \bibnamefont{and}
  \bibinfo{author}{\bibfnamefont{G.}~\bibnamefont{Misguich}},
  \emph{\bibinfo{title}{High Magnetic Fields}} (\bibinfo{publisher}{Springer},
  \bibinfo{address}{Berlin}, \bibinfo{year}{2002}), pp.
  \bibinfo{pages}{161--190}.

\bibitem[{\citenamefont{Antropov et~al.}(1997)\citenamefont{Antropov,
  Tretyakov, and Harmon}}]{ATH:JAP97}
\bibinfo{author}{\bibfnamefont{V.~P.} \bibnamefont{Antropov}},
  \bibinfo{author}{\bibfnamefont{S.~V.} \bibnamefont{Tretyakov}},
  \bibnamefont{and} \bibinfo{author}{\bibfnamefont{B.~N.}
  \bibnamefont{Harmon}}, \bibinfo{journal}{J. Appl. Phys.}
  \textbf{\bibinfo{volume}{81}}, \bibinfo{pages}{3961} (\bibinfo{year}{1997}).

\bibitem[{\citenamefont{Schmidt and Luban}(2003)}]{ScL:JPA03}
\bibinfo{author}{\bibfnamefont{H.-J.} \bibnamefont{Schmidt}} \bibnamefont{and}
  \bibinfo{author}{\bibfnamefont{M.}~\bibnamefont{Luban}}, \bibinfo{journal}{J.
  Phys. A: Math. Gen.} \textbf{\bibinfo{volume}{36}}, \bibinfo{pages}{6351}
  (\bibinfo{year}{2003}).

\bibitem[{\citenamefont{Schmidt et~al.}()\citenamefont{Schmidt, Schnack,
  Schr{\"o}der, and Luban}}]{SSS:05}
\bibinfo{author}{\bibfnamefont{H.-J.} \bibnamefont{Schmidt}},
  \bibinfo{author}{\bibfnamefont{J.}~\bibnamefont{Schnack}},
  \bibinfo{author}{\bibfnamefont{C.}~\bibnamefont{Schr{\"o}der}},
  \bibnamefont{and} \bibinfo{author}{\bibfnamefont{M.}~\bibnamefont{Luban}},
  \bibinfo{note}{unpublished}.

\bibitem[{\citenamefont{Brown et~al.}(2000)\citenamefont{Brown, Novotny, and
  Rikvold}}]{BNR:JAP00}
\bibinfo{author}{\bibfnamefont{G.}~\bibnamefont{Brown}},
  \bibinfo{author}{\bibfnamefont{M.~A.} \bibnamefont{Novotny}},
  \bibnamefont{and} \bibinfo{author}{\bibfnamefont{P.~A.}
  \bibnamefont{Rikvold}}, \bibinfo{journal}{J. Appl. Phys.}
  \textbf{\bibinfo{volume}{87}}, \bibinfo{pages}{4792} (\bibinfo{year}{2000}).

\bibitem[{\citenamefont{Derzhko and Richter}(2004)}]{DeR:PRB04}
\bibinfo{author}{\bibfnamefont{O.}~\bibnamefont{Derzhko}} \bibnamefont{and}
  \bibinfo{author}{\bibfnamefont{J.}~\bibnamefont{Richter}},
  \bibinfo{journal}{Phys. Rev. B} \textbf{\bibinfo{volume}{70}},
  \bibinfo{pages}{104415} (\bibinfo{year}{2004}).

\bibitem[{\citenamefont{Zhitomirsky and Honecker}(2004)}]{ZhH:JSM04}
\bibinfo{author}{\bibfnamefont{M.~E.} \bibnamefont{Zhitomirsky}}
  \bibnamefont{and} \bibinfo{author}{\bibfnamefont{A.}~\bibnamefont{Honecker}},
  \bibinfo{journal}{J. Stat. Mech.: Theor. Exp.} p. \bibinfo{pages}{P07012}
  (\bibinfo{year}{2004}).

\bibitem[{\citenamefont{Blake et~al.}(1997)\citenamefont{Blake, Gould, Grant,
  Milne, Parsons, and Winpenny}}]{BGG:JCSDT97}
\bibinfo{author}{\bibfnamefont{A.~J.} \bibnamefont{Blake}},
  \bibinfo{author}{\bibfnamefont{R.~O.} \bibnamefont{Gould}},
  \bibinfo{author}{\bibfnamefont{C.~M.} \bibnamefont{Grant}},
  \bibinfo{author}{\bibfnamefont{P.~E.~Y.} \bibnamefont{Milne}},
  \bibinfo{author}{\bibfnamefont{S.}~\bibnamefont{Parsons}}, \bibnamefont{and}
  \bibinfo{author}{\bibfnamefont{R.~E.~P.} \bibnamefont{Winpenny}},
  \bibinfo{journal}{J. Chem. Soc.-Dalton Trans.} pp. \bibinfo{pages}{485--495}
  (\bibinfo{year}{1997}).

\bibitem[{\citenamefont{Brechin et~al.}(1997)\citenamefont{Brechin, Graham,
  Harris, Parsons, and Winpenny}}]{BGH:JCSDT97}
\bibinfo{author}{\bibfnamefont{E.~K.} \bibnamefont{Brechin}},
  \bibinfo{author}{\bibfnamefont{A.}~\bibnamefont{Graham}},
  \bibinfo{author}{\bibfnamefont{S.~G.} \bibnamefont{Harris}},
  \bibinfo{author}{\bibfnamefont{S.}~\bibnamefont{Parsons}}, \bibnamefont{and}
  \bibinfo{author}{\bibfnamefont{R.~E.~P.} \bibnamefont{Winpenny}},
  \bibinfo{journal}{J. Chem. Soc.-Dalton Trans.} pp.
  \bibinfo{pages}{3405--3406} (\bibinfo{year}{1997}).

\end{thebibliography}

\end{document}